# n-type chalcogenides by ion implantation

Mark A. Hughes[1], Yanina Fedorenko[1], Behrad Gholipour[2], Jin Yao[2], Tae-Hoon Lee[3], Russell M. Gwilliam[1], Kevin P. Homewood[1], Steven Hinder[4], Daniel W. Hewak[2], Stephen R. Elliott[3] & Richard J. Curry[1]

Carrier-type reversal to enable the formation of semiconductor p-n junctions is a prerequisite for many electronic applications. Chalcogenide glasses are p-type semiconductors and their applications have been limited by the extraordinary difficulty in obtaining n-type conductivity. The ability to form chalcogenide glass p-n junctions could improve the performance of phase-change memory and thermoelectric devices and allow the direct electronic control of nonlinear optical devices. Previously, carrier-type reversal has been restricted to the GeCh (Ch = S, Se, Te) family of glasses, with very high Bi or Pb 'doping' concentrations (~5–11 at.%), incorporated during high-temperature glass melting. Here we report the first n-type doping of chalcogenide glasses by ion implantation of Bi into GeTe and GaLaSO amorphous films, demonstrating rectification and photocurrent in a Bi-implanted GaLaSO device. The electrical doping effect of Bi is observed at a 100 times lower concentration than for Bi melt-doped GeCh glasses.

[1] Department of Electronic Engineering, Advanced Technology Institute, University of Surrey, Guildford GU2 7XH, UK. [2] Optoelectronics Research Centre, University of Southampton, Southampton SO17 1BJ, UK. [3] Department of Chemistry, University of Cambridge, Lensfield Road, Cambridge CB2 1EW, UK. [4] The Surface Analysis Laboratory, Department of Mechanical Engineering Sciences, University of Surrey, Guildford GU2 7XH, UK. Correspondence and requests for materials should be addressed to M.A.H. (email: m.a.hughes@surrey.ac.uk) or D.W.H. (email: dh@orc.soton.ac.uk) or to S.R.E. (email: sre1@cam.ac.uk) or to R.J.C. (email: r.j.curry@surrey.ac.uk).

Chalcogenide glass alloys are of significant technological importance for nonlinear optics[1], infrared (IR) optics[2] and phase-change memory (PCM)[3]; however, electronic applications are limited as a consequence of their almost universally unmodifiable p-type conductivity[4]. The melt-doping technique previously used to obtain carrier-type reversal (CTR) in chalcogenide glass alloys is incompatible with the tooling used in integrated circuit (IC) fabrication; ion implantation, however, is integral to current IC fabrication. Therefore, CTR by ion implantation in chalcogenide glasses could open up a new branch of electronics based on these materials. Amorphous Se can be doped n-type with alkaline elements, which has allowed high-performance X-ray detectors to be fabricated[5]; however, amorphous Se is limited in its PCM, nonlinear and IR applications. GaLaSO glass is a high-performance PCM material that, compared with standard GeSbTe PCM materials, offers significantly higher thermal stability, and therefore potentially improved endurance, along with more than order of magnitude lower set and reset currents[6,7]. GeTe is an ingredient of standard GeSbTe PCM materials and is itself a useful PCM material, having a higher crystallization temperature and therefore the ability to operate at higher temperatures. A significant issue with PCM is the problem of leakage paths, which means that significant current can pass through unselected memory elements[8]. A solution is to insert a rectifying p-n junction in series with each memory element; p-n junctions in chalcogenide glasses could also lead to the development of novel photodetectors, light-emitting diodes, optical amplifiers and lasers. The injection of carriers into a p-n junction can modulate the nonlinear properties of the semiconductor used[9]; therefore, chalcogenide p-n junctions could be used for direct electronic control of optical ultrafast nonlinear devices such as demultiplexers, wavelength converters and optical Kerr shutters. Among the chalcogenide glasses, GaLaSO is particularly notable with respect to optical nonlinear devices as it has the highest nonlinear figure of merit (FOM) of any glass reported to date[10], with FOM $= n_2/2\beta\lambda$, where $\lambda$ is the wavelength, $n_2$ is the real part of the nonlinear refractive index and $\beta$ is the two-photon absorption coefficient. Ultraviolet-written waveguide[11] and microsphere[12] lasers have also been demonstrated in rare-earth-doped GaLaSO. Chalcogens are important components of thermoelectric materials, particularly when alloyed with Bi or Pb; therefore, Bi-modified chalcogenide glasses could have applications in novel thermoelectric devices.

In this work, we present, to the best of our knowledge, the first demonstration of CTR by ion implantation of a chalcogenide glass, the first demonstration of CTR by Bi doping in a non-GeCh glass and the lowest doping concentration (0.6 at.%) for CTR in any chalcogenide glass alloy. The two alloys we used represent different classes of chalcogenide glasses: GaLaSO is highly resistive but has excellent optical properties, whereas GeTe is conductive but has poor optical properties.

## Results

**Material characterization and simulation.** Figure 1a shows the X-ray photoelectron spectroscopy (XPS) spectrum of Bi-implanted GaLaSO and GeTe, along with those for Bi metal and Bi melt-doped GeSe and silicate glass. The binding energy of Bi is determined mainly by its oxidation state, and tends to increase with increasing oxidation state. The broad $4f_{7/2}$ level of Bi melt-doped silicate at 160.0 eV contains components of $Bi^{5+}$ at 160.6 eV and $Bi^{4+}$ at 159.8 eV[13]; $Bi_2O_3$ has a binding energy of 158.9 eV corresponding to $Bi^{3+}$ (ref. 14). The peak for Bi metal at 156.9 eV corresponds to neutral Bi. The broad $4f_{7/2}$ level of Bi-implanted GaLaSO can be deconvolved, see Supplementary Fig. 1, into peaks at 157.1 eV, which corresponds to neutral Bi, and 158.1 eV, which corresponds to a positively charged species with an oxidation state $<3+$, although $Bi^+$ and $Bi^{2+}$ cannot be distinguished at this binding energy. Supplementary Figure 2a,b indicates that X-ray exposure does not induce a valence state change in Bi during measurement. Bi-implanted GeTe also has a single peak 157.9 eV corresponding to $Bi^+$ or $Bi^{2+}$ only.

Figure 1b shows the XPS depth profile of a $1\times10^{16}$ ions cm$^{-2}$ Bi-implanted GaLaSO thin film; the Bi $4f_{7/2}$ peak from the implantation can be seen appearing below the surface of the film. Figure 1c shows the Bi concentration, calculated by fitting the photoelectron spectra, as a function of etching time. The concentration peaks at 0.6 at.% at an etch time of 7,000 s; the film composition was $Ga_{25}La_{32}S_{28}O_{15}$ and was constant throughout the etch. Rutherford backscattering measurements indicate a similar film composition and resolved a single buried layer of Bi in the film with a concentration of 0.32 at.% to a depth of 84 nm (errors indicated in Fig. 1c), which is consistent with the XPS measurement.

The depth profile of the $2\times10^{16}$ ions cm$^{-2}$ Bi-implanted GeTe film in Fig. 2a shows that the peak Bi concentration was 1.4 at.%. It also shows that the composition was $\sim Ge_{33}Te_{33}O_{33}$ at the surface, $Ge_{15}Te_{60}O_{15}$ at the centre of the film and $Ge_{15}Te_{85}$ at the interface with the substrate. The O profile shows that O does not reach the film/substrate interface, which indicates that the film has oxidized due to exposure to atmospheric $O_2$. Rutherford backscattering and particle-induced X-ray emission measurements of this film taken 1 week after deposition indicated that the majority of oxidation had already occurred before ion implantation. The composition of the sputter target was $Ge_{50}Te_{50}$, indicating that Te is preferentially sputtered during film deposition. Sputter markers indicated a $\sim 40$-nm sputter depth after $2\times10^{16}$ ions cm$^{-2}$ Bi implantation. In contrast to GeTe and many other chalcogenide films, we found that GaLaSO is remarkably stable against atmospheric oxidation and also against sputtering during implantation. As shown in the depth profile of $1\times10^{16}$ ions cm$^{-2}$ Bi-implanted GaLaSO in Fig. 2b, the O profile is flat, indicating that oxidation due to atmospheric $O_2$ did not occur, since this would increase the O content towards the surface of the film. The flat composition profile through the Bi implant indicates that the penetration depth of collision cascades created by the implantation was low enough not to affect the composition. The flat profile also indicates that sputter profiling during the XPS measurement, where components are preferentially sputtered or driven into the film, creating a false profile, is unlikely because, to get a flat profile when sputter profiling is occurring, the actual profile would have to, by chance, be the inverse of the sputter-induced profile. The composition of the sputter target was $Ga_{28}La_{12}S_{18}O_{42}$, indicating that La is preferentially sputtered and O and S are unfavourably sputtered during film deposition. Sputter markers indicated no sputtering of the GaLaSO film after implantation, even with a dose of $2\times10^{16}$ ions cm$^{-2}$.

Extended X-ray absorption fine structure spectroscopy of bulk GaLaS glasses with compositions spanning the glass-forming region indicated that $GaS_4$ tetrahedra are the main glass former units for all compositions[15]. XPS measurements of Ga and O, within and below the Bi-implanted region of GaLaSO, can be seen in Fig. 3a,b, respectively. In the implanted region, there is a decrease in the high binding energy region of the Ga peak. Analysis of this Ga peak in Supplementary Fig. 3 indicates the presence of homopolar Ga–Ga bonds, and that Ga is primarily coordinated with S. It also shows that implantation decreases the presence of Ga species in an environment similar to $\beta$-$Ga_2O_3$ and increases the number of $GaS_4$ tetrahedra. Raman spectra in Supplementary Fig. 4 also indicate that $GaS_4$ tetrahedra are formed after Bi implantation. The formation of GaS$_4$ tetrahedra is feasible if we assume that the tetrahedra are arranged in a similar manner to Ga$_2$S$_3$ (where S/Ga = 1.5). If we take into account the error in the composition measurement, the S/Ga ratio of our film is ~0.9–1.4. If we also take into account the existence of Ga–Ga and Ga–O bonds in GaLaSO, the S/Ga ratio for Ga coordinated with S could be similar to that of Ga$_2$S$_3$.

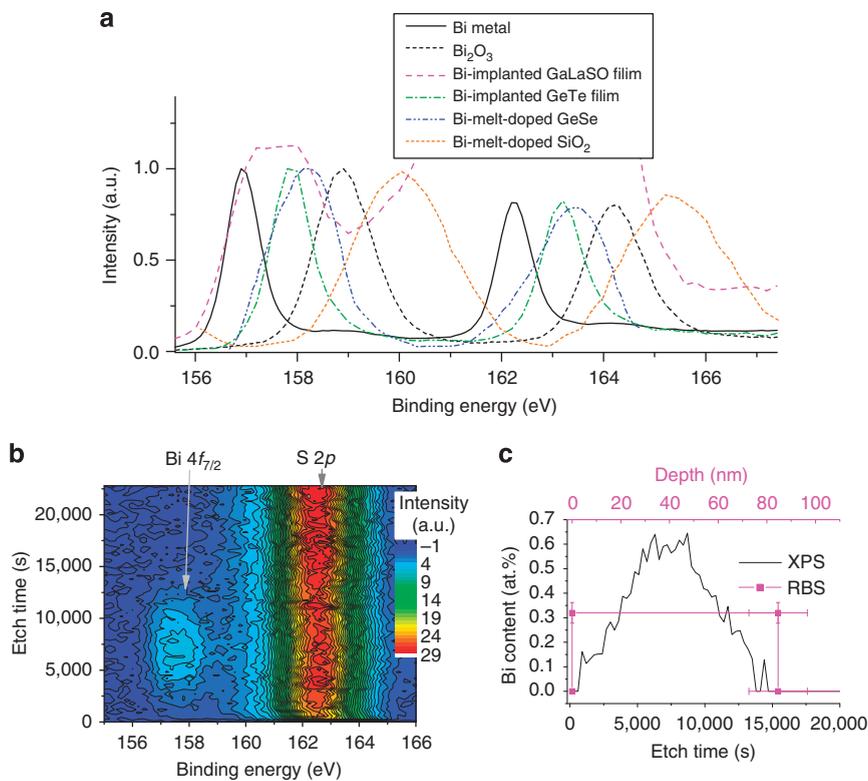

**Figure 1 | XPS measurements of Bi properties in various host environments.** (**a**) XPS spectrum of Bi from n-type $1 \times 10^{16}$ ions cm$^{-2}$ Bi-implanted GaLaSO, $2 \times 10^{16}$ ions cm$^{-2}$ Bi-implanted GeTe (taken at the centre of the Bi implants), Bi metal and Bi$_2$O$_3$. Also, the XPS spectrum for melt-doped Si$_{28}$O$_{66}$Al$_4$Bi$_2$ glass after Fujimoto[13], and melt-doped n-type Ge$_{20}$Se$_{70}$Bi$_{10}$ after Kumar[35], are shown. The two low and high binding energy peaks, which should have equal spacing, originate from the Bi 4$f_{7/2}$ and 4$f_{5/2}$ levels, respectively. In GaLaSO, the Bi 4$f_{5/2}$ peak is obscured by the S 2$p$ peak, which can be observed throughout the entire etch; because of the equal spacing of the peaks, only a single peak is necessary to deduce the binding energy. (**b**) XPS depth profile, representing XPS spectra taken during a sequence of etch cycles, of a $1 \times 10^{16}$ ions cm$^{-2}$ Bi-implanted 200-nm-thick GaLaSO film; the Bi 4$f_{7/2}$ peak can be seen appearing between etch times of 5,000 and 10,000 s, the scale is counts s$^{-1}$. (**c**) Bi concentration in a $1 \times 10^{16}$ ions cm$^{-2}$ Bi-implanted GaLaSO thin film, measured by XPS, as a function of etch time and calculated by fitting the Bi 4$f_{7/2}$ peak at 157.5 eV. The Bi depth profile measured by Rutherford backscattering (RBS), limited to a single layer resolution, is also shown with a scale in nm. RBS errors show one s.d. obtained from counting statistics and the signal-to-noise ratio.

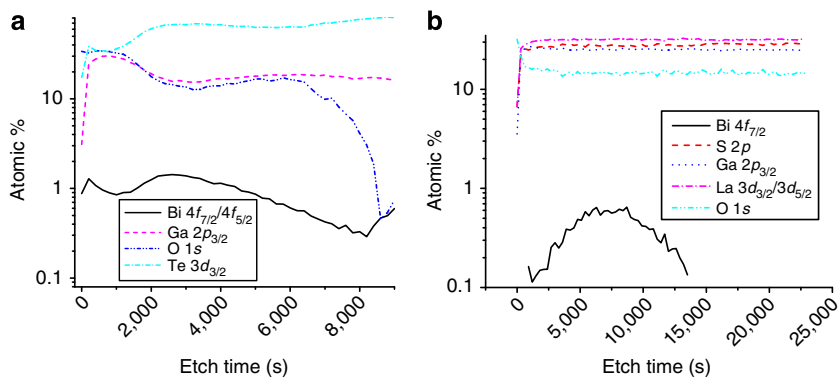

**Figure 2 | XPS measurement of GeTe and GaLaSO film stoichiometry.** (**a**) Depth profile of a $2 \times 10^{16}$ ions cm$^{-2}$ Bi-implanted 100-nm-thick GeTe film measured by XPS; the end of the etch is at the film/substrate interface. Compositions were calculated by fitting the indicated energy level(s) for each component. (**b**) Depth profile of a $1 \times 10^{16}$ ions cm$^{-2}$ Bi-implanted 200-nm-thick GaLaSO film measured by XPS. Compositions were calculated by fitting the indicated energy level(s) for each component. The composition was Ga$_{25}$La$_{32}$S$_{28}$O$_{15}$ and was constant throughout the etch.

Figure 3b shows that there is an increase in the high binding energy region of O in the implanted region. Analysis of this O peak in Supplementary Fig. 5 indicates that implantation leads to an increase in an O species in an environment similar to that in α-Ga$_2$O$_3$. Supplementary Fig. 6 shows that the differences observed in these spectra between the implanted and unimplanted regions were significantly greater than that between two spectra from the same region. Because of the low doping concentration of Bi, it is unlikely that the changes in the XPS spectra are due to the presence of Bi itself, but more likely due to the structural rearrangement caused by the Bi implantation.

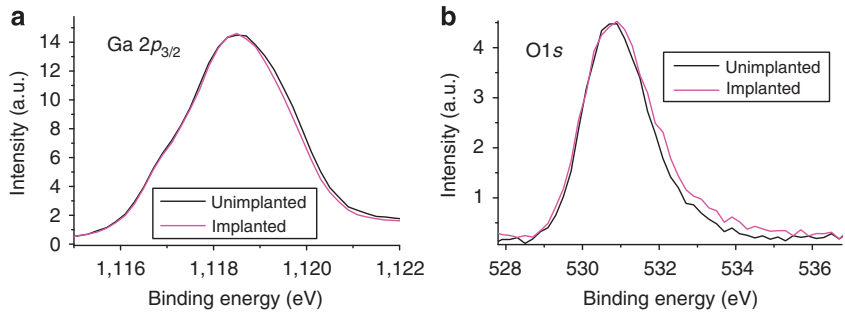

**Figure 3 | XPS measurements of Ga and O binding energy in GaLaSO films.** (**a**) XPS spectra of the Ga $2p_{3/2}$ level from the unimplanted region below the Bi implant and from within the Bi-implanted region in a $1 \times 10^{16}$ ions cm$^{-2}$ Bi-implanted 200-nm-thick GaLaSO film. (**b**) XPS spectra of O 1s from the unimplanted region below the Bi implant and from within the Bi-implanted region.

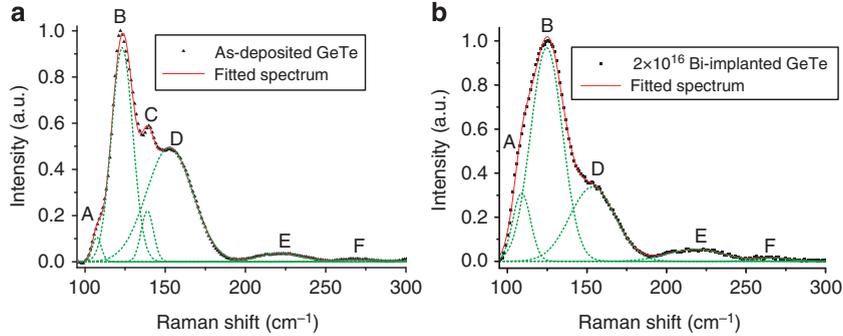

**Figure 4 | Raman spectroscopy of GeTe films.** Raman spectra, excited at 782 nm and deconvolved into fitted Gaussians, of 100-nm-thick GeTe films on fused silica substrates. (**a**) As-deposited film. (**b**) Bi-implanted film ($2 \times 10^{16}$ ions cm$^{-2}$).

As shown in Supplementary Fig. 7, there were no differences in the XPS spectra of the implanted and unimplanted regions for La, and analysis indicates that La is primarily coordinated with O. The XPS of S in the implanted and unimplanted regions could not be compared because of the overlapping of the Bi $4f_{5/2}$ peak; however, analysis of the unimplanted region in Supplementary Fig. 8 confirms that La and Ga are primarily coordinated with O and S, respectively, and indicates the presence of homopolar S–S bonds. The absorption spectra in Supplementary Fig. 9 indicate the formation of homopolar S–S bonds following implantation. XPS spectral analysis is not presented for GeTe because the compositional variation through the film makes the interpretation of any differences in the implanted and unimplanted regions less clear.

We therefore propose that Bi implantation leads to the formation of GaS$_4$ tetrahedra and homopolar S–S bonds while increasing and decreasing α- and β-Ga$_2$O$_3$-like phases, respectively; the balance of this bond breaking and formation would be made up by rearrangement of the GaLaSO network.

The Raman spectra of unimplanted and Bi-implanted GeTe films, shown in Fig. 4a,b, respectively, indicate significant structural modification after implantation. Tetrahedral species of the type GeTe$_{4-n}$Ge$_n$ ($n = 0, 1, 2, 3, 4$) are the main building blocks of amorphous GeTe[16]. The Raman spectrum of unimplanted GeTe was typical of that for sputtered amorphous GeTe films[17], with a deconvolution into Gaussians revealing six bands at 107, 122, 139, 153, 222 and 269 cm$^{-1}$ (A-F, respectively). Band A has been attributed[16,17] to corner-sharing GeTe$_4$ and GeTe$_3$Ge tetrahedra; the relative intensity of this band increased by ~200% in the Bi-implanted film. Band B has been attributed to GeTe$_2$Ge$_2$ and GeTeGe$_3$ tetrahedra. The bandwidths of bands A and B increase after implantation, indicating an increased structural disorder, such as a greater variation in bond lengths or angles, in the tetrahedra attributed to them. If we assume a 50/50 mixture of each of the two tetrahedra attributed to bands A and B, then the 200% increase in band A corresponds to a 38% increase in Ge–Te bonds. Band C is attributed to oxidation of the film, and is not visible following implantation. Band D has been attributed to disordered Te chains and is around 30% weaker after implantation. Band E is attributed to antisymmetric stretching modes of the tetrahedra and is 50% stronger in the implanted film. Band F is attributed to Ge–Ge bonds and is unchanged after implantation. Our Raman measurements indicate a 38% increase in Ge–Te bonds, and a 30% reduction in a band associated with Te chains with Bi implanted to a peak concentration of 1.4 at.%. *Ab initio* modelling, Fig. 5, indicates a 4% increase in Ge–Te bonds and a 4% decrease in Te–Te wrong bonds with 2 at% Bi doping. The *ab initio* model is for a quenched system, so it corresponds more to a melt-doped glass than an ion-implanted glass. The *ab initio* model indicates a similar structural effect to the implanted GeTe; however, the effect is an order of magnitude stronger in the implanted GeTe with similar Bi concentrations. This suggests that implantation of Bi accentuates its chemical effect compared with when it is melt-doped; this could explain why we are able to observe a much lower Bi concentration for CTR in Bi-implanted GeTe compared with Bi melt-doped GeTe (Fig. 6d). We were unable to successfully perform *ab initio* modelling of GaLaSO because of the complexity of the La electronic structure.

**Electrical characterization.** Figure 6a shows the electrical conductivity as a function of inverse temperature of GaLaSO thin films for various Bi doses. The temperature dependence of the conductivity was the same for the undoped materials and all doses below $1\times10^{15}$ ions cm$^{-2}$. However, at a dose of $3\times10^{15}$ ions cm$^{-2}$, the conductivity increased and the activation energy decreased, this trend continues as the dose is increased. The conductivity diverged slightly from its Arrhenius behaviour at lower temperatures for doses between $3\times10^{15}$ and $2\times10^{16}$ ions cm$^{-2}$, indicating the presence of hopping conduction. Doses below $1\times10^{15}$ ions cm$^{-2}$ showed no evidence for hopping conduction; however, in the temperature regime where hopping conduction occurred at higher doses, the conductivity could not be measured because the conductance was below the system detection limit. Figure 6b shows the conductivity as a function of Bi dose, for implanted GaLaSO and GeTe films, and as a function of Bi concentration for bulk Bi melt-doped GeTe and GeSe glasses. Bi melt-doped GeSe has a threshold dependence of conductivity on the Bi concentration where Bi initially has little electrical doping effect, but then causes a sudden increase in conductivity over several orders of magnitude; this increase in conductivity is associated with CTR[18]. Analogous behaviour can be seen in Bi-implanted GaLaSO; however, the onset of the electrical doping effect of Bi occurs at two orders of magnitude lower concentration than in Bi melt-doped GeSe. We also implanted a GaLaSO film with $3\times10^{16}$ ions cm$^{-2}$ of Xe and found no increase in conductivity compared with the unimplanted film, indicating that the increase in conductivity is due to Bi itself, rather than the implantation damage it causes. Bi has far less effect on the conductivity in melt-doped GeTe; the effect is stronger in Bi-implanted GeTe, but there is still no threshold dependence. Figure 6c shows the activation energy for electrical conductivity as a function of Bi content for Bi-implanted GaLaSO and GeTe films and Bi melt-doped bulk GeTe and GeSe. In Bi melt-doped GeSe, a sudden decrease in the activation energy indicates unpinning of the Fermi level, which accompanies CTR[18]. GaLaSO also shows a sudden decrease in the activation energy again appearing at a Bi concentration two orders of magnitude lower than in Bi melt-doped GeSe. In contrast, the activation energy changes gradually for both Bi-implanted and melt-doped GeTe.

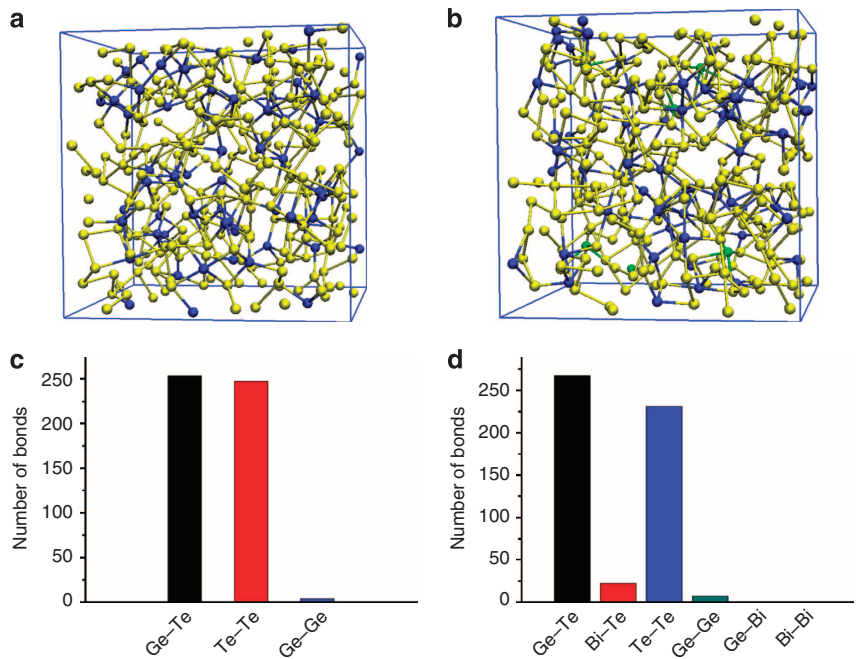

**Figure 5 | Results of *ab initio* structural modelling.** *Ab initio* structural model of $Ge_{20}Te_{80}$ (**a**) and $Ge_{20}Te_{78}Bi_2$ (**b**); blue: Ge, yellow: Te, green: Bi. *Ab initio* modelling result of the short-range chemical order in $Ge_{20}Te_{80}$ (**c**) and $Ge_{20}Te_{78}Bi_2$ (**d**). The modelling indicates that Bi doping reduces wrong bonds (mostly Te–Te) by ~4% in the amorphous structure and increases Ge–Te bonds by ~4%.

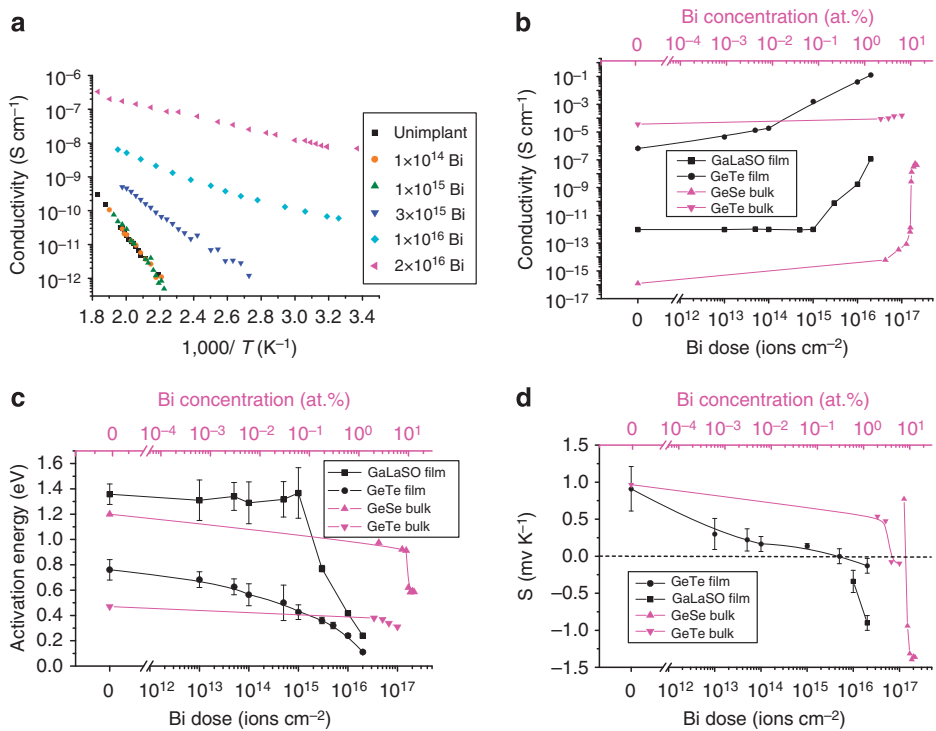

**Figure 6 | Electrical properties of Bi-implanted chalcogenide films.** (**a**) Electrical conductivity as a function of temperature for a 100-nm-thick GaLaSO film, implanted with Bi at various doses in unit ions cm$^{-2}$. (**b**) Electrical conductivity as a function of Bi concentration for 100-nm-thick Bi-implanted GaLaSO and GeTe films, Bi melt-doped GeTe after Bhatia[36] and GeSe after Tohge[18]. The conductivity of GaLaSO was measured at 180 °C, all other samples were measured at room temperature. (**c**) Activation energy for the electrical conductivity as a function of Bi concentration for 100-nm-thick Bi-implanted GaLaSO and GeTe films, Bi melt-doped GeTe after Bhatia[36] and GeSe after Tohge[18]. (**d**) Seebeck coefficient as a function of Bi concentration for 100-nm-thick Bi-implanted GaLaSO and GeTe films, Bi melt-doped GeTe after Bhatia[36] and GeSe after Tohge[18]. The Bi implant dose and Bi concentration axis were scaled to each other using the XPS measurement of a peak Bi concentration of 0.6 at.% at a dose of $1 \times 10^{16}$ ions cm$^{-2}$. The Rutherford backscattering measurement could not resolve a peak concentration, so was not used to scale the axis. All error bars represent one s.d. of experimental variation and lines are provided as a guide to the eye.

Figure 6d shows the Seebeck coefficient (S) as a function of Bi content for Bi-implanted GaLaSO and GeTe films and Bi melt-doped bulk GeTe and GeSe. The $1\times10^{16}$ and $2\times10^{16}$ ions cm$^{-2}$-implanted GaLaSO samples had a negative S, indicating n-type conductivity. The concentration of 0.6 at.% Bi at which we observe CTR in GaLaSO is at an order of magnitude lower concentration than has been reported in melt-doped glasses. If the Bi was distributed homogeneously through the film, as could be achieved by using a chain of Bi implants at different energies, CTR could occur at a significantly lower Bi concentration. Samples with lower doses were too resistive to measure S; however, all undoped chalcogenides reported in the literature are p-type, so it is reasonable to assume that unimplanted GaLaSO is also p-type. Unimplanted GeTe is five orders of magnitude more conductive than GaLaSO at 180 °C; consequently, we were able to measure S in GeTe as it changed from being positive in the unimplanted film to being negative in the $2\times10^{16}$ ions cm$^{-2}$ implanted film.

Bulk undoped and 0.4 at.% Bi melt-doped GaLaSO had the same activation energy as for the unimplanted GaLaSO film, and there was no change in activation energy or conductivity with Bi doping. This indicated that the low-concentration electrical doping effect of Bi on GaLaSO requires that it is implanted, rather than melt doped. Melt doping of Bi in glasses is associated with a higher Bi oxidation state; this oxidation state results from the redox equilibrium reached between Bi and its host during the melt-quench process. Ion implantation circumvents this redox equilibrium, and implanted ions are more likely to sit in interstitial sites, whereas melt-doped ions tend to sit on lattice sites. The unyielding p-type conductivity of chalcogenide glasses originates from charged defects known as valence-alternation pairs (VAPs): triply coordinated, positively charged chalcogens ($Ch_3^+$) and singly coordinated, negatively charged chalcogens ($Ch_1^-$). Excitation of $Ch_3^+$ leads to its decay into $Ch_1^0$ resulting in a recombed electron and an uncombined hole, whereas excitation of $Ch_1^-$ leads to a recombined hole and electron[19]. The mechanism of CTR in melt-doped Bi:GeCh glasses is disputed. Phillips proposed that they contain $Bi_2Ch_3$ clusters $\sim 3.5$ nm in diameter with a tetradymite structure, embedded in a GeCh matrix. CTR may be effected by excess chalcogen atoms on the cluster surface, which help to reduce the number of dangling bonds[20]. Our XPS measurements, which show neutral Bi and $Bi^+$ or $Bi^{2+}$ in implanted GaLaSO, are inconsistent with the formation of $Bi_2Ch_3$ clusters. The Phillips model lends itself well to explaining the threshold increase in conductivity in Bi:GeCh glasses as a percolation threshold, in which regions of high-conductivity $Bi_2Ch_3$ in a low-conductivity GeCh glass matrix form an infinite cluster above a critical volume fraction, resulting in a sudden increase in conductivity. Geometric simulations of disordered semiconductors show that the percolation threshold for hopping conductivity occurs at a volume fraction of 0.219 (ref. 21), which is in good agreement with the $\sim 0.22$ volume fraction of $Bi_2Ch_3$ at which a percolation threshold for conductivity is observed in melt-doped $Ge_{20}Ch_{80-x}Bi_x$ (Ch = S, Se) glasses[22]. We observe a percolation threshold at two orders of magnitude lower Bi concentration in Bi-implanted GaLaSO. For a percolation threshold to occur at such a low doping concentration, the dopant must have a large aspect ratio, as occurs in carbon nanotube-doped polymers[23], or the volume fraction of the dopant must be much larger than would be expected for its concentration. This indicates that the threshold increase in conductivity in Bi-implanted GaLaSO does not result from a percolation threshold. In addition, Bi-implanted and melt-doped GeTe do not display threshold behaviour.

We also observed a threshold for near IR (NIR) photoluminescence (PL) bands in Bi-implanted GaLaSO excited at 514 nm[24]; at low doses, only a 700-nm band was observed, and above $1\times10^{15}$ ions cm$^{-2}$ a NIR band was observed. The 700-nm band is often associated with Bi-doped oxide glasses, whereas the NIR band is observed in Bi-doped chalcogenides. We therefore propose that the implanted Bi preferentially associates with an oxide site that is electrically inactive and generates 700 nm PL; at higher doses, the oxide sites are filled and Bi associates with a sulphide site that is related to electrical doping and generates NIR PL. From extended X-ray absorption fine structure spectroscopy measurements of $Ge_{20}S_{80-x}Bi_x$ ($x=3-10$), Elliott et al.[25] argued that Bi is only threefold coordinated, asserted that the glasses were homogeneous without $Bi_2Ch_3$ clusters and that the mechanism of CTR is due to the presence of charged Bi atoms, which suppress the concentration of positively charged VAPs at the expense of negatively charged VAPs and, as a result, the electron concentration is increased. This is accompanied by a shift from purely covalent to partially ionic bonding between Bi and chalcogen atoms, with the most favourable configuration being $Bi_3^+$ (ref. 26). Other mechanisms invoking unpinning of the Fermi level by centres including $Bi_4^+$ (ref. 27) or $Bi_2^-$ (ref. 28) have been proposed. Our results lend themselves better to a model in line with that of Elliott et al. in which Bi is homogeneous and initiates electronic doping by affecting the balance of VAPs. Our XPS measurements show that n-type Bi-implanted GeTe and GaLaSO both contain a positively charged Bi species ($Bi^+$ or $Bi^{2+}$), and GaLaSO also contains neutral Bi. Since $Bi^+$ or $Bi^{2+}$ is the common species, it, rather than neutral Bi, must be responsible for CTR.

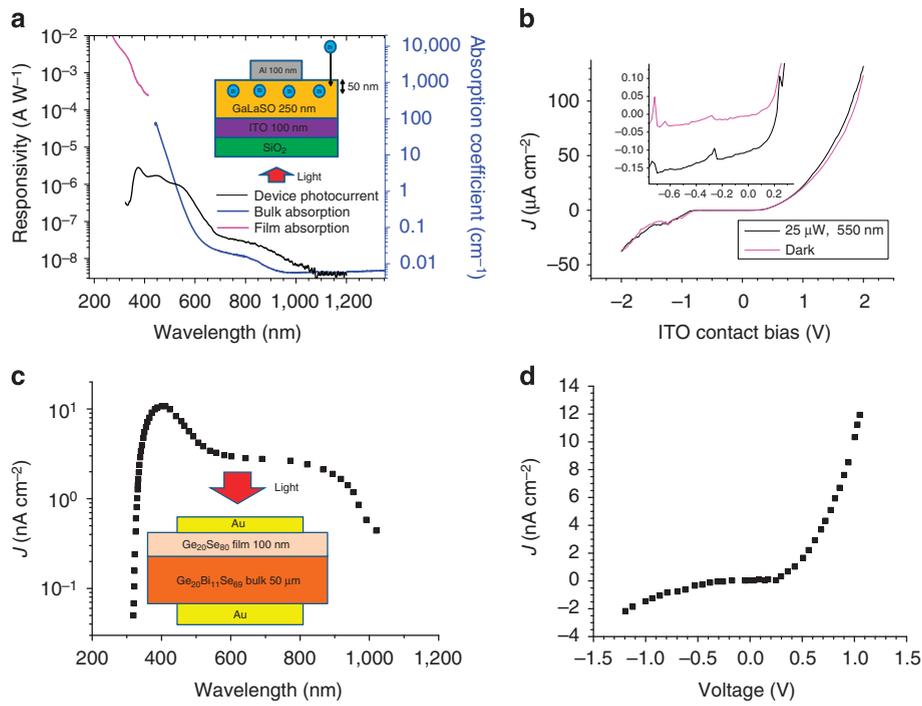

**Figure 7 | GaLaSO p-n junction device characterization and comparison.** (**a**) Responsivity (photocurrent/illumination power) as a function of wavelength for a Bi-implanted GaLaSO device. The device structure is illustrated in the inset and the Bi dose was $1 \times 10^{16}$ ions cm$^{-2}$. The absorption spectra of bulk GaLaSO and a 100-nm-thick unimplanted film are also shown. (**b**) Dark and illuminated J–V characteristics of the Bi-implanted GaLaSO device. The inset shows a close-up of the low-current region. (**c**) Photocurrent spectrum from a Bi melt-doped GeSe device after Tohge[30], consisting of an n-type bulk Ge$_{20}$Bi$_{11}$Se$_{69}$ layer and a p-type evaporated Ge$_{20}$Se$_{80}$ layer. The device schematic is shown in the inset. (**d**) Dark J–V characteristic of the device shown in **c**.

We recently proposed that Bi-related CTR and PL are caused by the same active centre and reported PL from Bi-implanted GaLaSO[24]. In this work, we observed Bi-related CTR from Bi-implanted GaLaSO that contains neutral Bi and Bi$^+$ or Bi$^{2+}$. PL is unlikely to originate from neutral Bi; although it can be observed in neutral metals[29], it is generally stronger when they are oxidized. In addition, we were unable to measure any PL from metallic Bi using the same system we used to measure PL in Bi-implanted GaLaSO. This implies that these disparate phenomena originate from the same Bi centre (Bi$^+$ or Bi$^{2+}$).

**Device characterization.** We fabricated a device with a structure as shown schematically in the inset of Fig. 7a. The reasoning behind this choice of device structure is that we expect the implanted region close to the surface to be n-type, whereas the lower portion of the film should be p-type and hence should set up a space–charge region. We were able to detect photocurrent from the device when it was unbiased, indicating the presence of such a space–charge region. In an unimplanted control device, no photocurrent could be measured. The device responsivity, $R(\lambda)$, was calculated from the photocurrent spectrum, $I_{pc}(\lambda)$, and the output power spectrum of the monochromator, $P_{opt}(\lambda)$, with $R(\lambda) = I_{pc}(\lambda)/P_{opt}(\lambda)$. The responsivity spectrum is a good match to the absorption spectrum of bulk GaLaSO, including the deep bandgap states in the ~550–1,100 nm region, as direct absorption into mid-gap states occurs. As the wavelength is reduced further (below 550 nm), the photocurrent is reduced in comparison with the absorption. The change in behaviour at this wavelength close to the bandgap indicates that alternative competing relaxation pathways exist for photoexcited charge carriers which do not lead to the generation of photocurrent. Nonetheless, some photoexcited charge carriers do relax into states from which a photocurrent can be extracted. Finally, at wavelengths below ~370 nm (3.35 eV), the photocurrent is reduced as the indium tin oxide (ITO) substrate attenuates the incident light.

The current density–voltage (J–V) characteristics of the device, inset of Fig. 7b, show that the device has some rectification behaviour and an open-circuit voltage of 0.2 V under 25 µW of illumination power (3.2 mW cm$^{-2}$ power density). The non-ideal rectifying characteristics could be due to current leakage across the junction. Figure 7c shows the photocurrent spectrum of a Bi melt-doped GeSe-based p-n junction device, after Tohge[30]. The spectral response of our GaLaSO device and the GeSe device are observed to be similar, as would be expected since GeSe has a similar bandgap to GaLaSO. However, there was no correction for, or mention of, the illumination power used when measuring the GeSe device, so the responsivity of the two devices cannot be compared. The dark J–V curve of the GeSe device is shown in Fig. 7d and takes the same form as that of the GaLaSO device, but over a narrower voltage range. The forward J at 1 V is approximately three orders of magnitude greater in our GaLaSO device. We attribute this to our significantly thinner structure, which would reduce overall resistance, and our use of a single chalcogenide layer, which would improve the interface between p and n regions.

In conclusion, Bi-implanted GaLaSO thin films display a threshold dependence of the electrical doping effect of Bi. This threshold occurs at two orders of magnitude lower Bi concentration than that in Bi melt-doped GeCh glasses. The Bi concentration at which this threshold occurred was too low for it to be caused by a percolation threshold. In Bi-implanted GeTe, there is no threshold dependence for the electrical doping effect of Bi. We report CTR in $1 \times 10^{16}$ and $2 \times 10^{16}$ ions cm$^{-2}$ Bi implanted in GaLaSO and GeTe chalcogenide films, respectively. This represents the first demonstration of CTR by ion implantation of chalcogenide glasses. CTR in Bi-implanted GaLaSO represents the first demonstration of CTR by Bi doping in a non-GeCh glass and the lowest doping concentration (0.6 at.%) for CTR in a chalcogenide glass alloy. XPS measurements indicate that the Bi species responsible for CTR in Bi-implanted GaLaSO and GeTe was $Bi^+$ or $Bi^{2+}$. Raman measurements of Bi-implanted GeTe and *ab initio* modelling of Bi-doped GeTe indicate that implantation accentuates the chemical effect of Bi on the structure of GeTe, which could explain why we are able to observe CTR at a much lower Bi concentration when Bi is implanted compared with when it is melt-doped. We fabricated a Bi-implanted GaLaSO p-n junction device that rectified and had a photocurrent spectrum that matched that of the bulk absorption spectrum. We also observed Bi-related PL and CTR from Bi-implanted GaLaSO, indicating that these disparate phenomena originate from the same Bi centre ($Bi^+$ or $Bi^{2+}$).

## Methods

**Chalcogenide film preparation and ion implantation.** A gallium lanthanum sulphur oxide (GLSO) sputtering target was prepared by mixing 70% gallium sulphide and 30% lanthanum oxide in a dry-nitrogen-purged glove box. The raw materials were melted for 24 h in dry argon, in 2 inch diameter vitreous carbon crucibles, annealed at the glass transition temperature and then sliced to form a 3-mm-thick sputter target. A $Ge_{50}Te_{50}$ sputtering target was purchased commercially from Testbourne. We sputtered 100-nm-thick films of GaLaSO and GeTe onto 15-µm-thick thermally oxidized $SiO_2$ on Si and fused silica substrates. Bismuth ions were implanted using a Danfysik ion implanter at an energy of 190 keV. During implantation, the samples were mounted on a carousel holder held at ambient temperature, and the beam current was kept below $1\,\mu A\,cm^{-2}$ to avoid beam heating of the samples. At a dose of $2\times 10^{16}$ ions cm$^{-2}$, sputter markers indicated no change in the film thickness of GaLaSO greater than the detection limit of 10 nm; for GeTe, sputter markers indicated ~40 nm of sputtering.

**Electrical and optical characterization.** Conductivity measurements were made by taking current–voltage (I–V) scans laterally across the films using a Keithley 4200 SCS setup incorporating remote preamplifiers within a Faraday cage[31]. For thermopower measurements, we heated below one electrical contact with a d.c. heating element and measured the temperature at each contact with a thermocouple. For device fabrication, a 250-nm film of GaLaSO was sputtered onto a substrate of 100-nm ITO on borosilicate glass. After implantation of Bi with a dose of $1\times 10^{16}$ ions cm$^{-2}$ and energy of 190 keV, a 100-nm-thick Al top contact with a diameter of 500 µm was sputtered. We also fabricated a control device with the same fabrication steps, but without a Bi implant. Absorption measurements were taken on a Varian Carey 5000 spectrophotometer. Raman measurements were performed on a Renishaw 2000 microRaman system.

**XPS analysis.** XPS analyses were performed on a ThermoFisher Scientific Theta Probe spectrometer using a monochromated Al Kα X-ray source ($h\nu = 1,486.6$ eV) with a spot radius of ~400 µm. High-resolution core-level spectra were acquired with a pass energy of 50 eV. All spectra were charge referenced against the C1s peak at 285 eV to correct for charging effects during acquisition. Quantitative surface chemical analyses were calculated from core-level spectra following the removal of the Shirley background and correction for the electron energy analyser transmission function. Depth profiles were acquired using 3 kV $Ar^+$ ions delivering ~1 µA of etch current. The $Ar^+$ beam was raster scanned over a $3\times 3\,mm^2$ area. A 60-s etch time was employed per etch level. The sample used for XPS was a $1\times 10^{16}$ ions cm$^{-2}$ Bi-implanted 250-nm-thick GaLaSO film.

**Device characterization.** Photocurrent measurements were taken by passing a 100-W halogen white-light source through a monochromator with 1,200 or 600 lines mm$^{-1}$ gratings. Monochromatic light was focused through the transparent substrate of the device and modulated with a mechanical chopper; the photocurrent was detected with a lock-in amplifier. The wavelength dependence of the output optical power from the monochromator was measured with an optical power meter.

**Structural simulations.** *Ab initio* molecular dynamics simulations were performed using the Vienna *ab initio* Simulation Package (VASP)[32]. The projector augmented wave method[33] with the Perdew–Burke–Ernzerhof exchange-correlation functional[34] was used. All the outer *s* and *p* electrons were treated as valence electrons. The plane-wave energy cutoff was 175 eV. The temperature was controlled by a Nose thermostat algorithm. The time step for molecular dynamic simulations was 4 fs. The models were simulated in cubic supercells with periodic boundary conditions. The atomic configurations were first mixed at 3,000 K, maintained at 1,000 K for tens of ps and then were quenched to generate amorphous models with a quench rate of $-15\,K\,ps^{-1}$. The model density used was $5.9\,g\,cm^{-3}$.

## Acknowledgements
This work was supported by the UK EPSRC grants EP/I018417/1, EP/I019065/1 and EP/I018050/1. We would like to thank Mr Chris Craig for fabricating the GaLaSO sputtering target and Professor John Watts for helpful discussions on XPS measurements.

## Author contributions
The concept was developed by D.W.H., S.R.E. and R.J.C. Experimental work was performed by M.A.H., Y.F., B.G., J.Y. and S.H. with input from R.M.G., K.P.H., D.W.H. and R.J.C. Modelling was performed by T.-H.L. with input from S.R.E. M.A.H. wrote the manuscript with editorial input from Y.F., D.W.H., S.R.E. and R.J.C. All authors contributed to analysing the results and commented on the paper.

## Additional information
**Supplementary Information** accompanies this paper at http://www.nature.com/naturecommunications

**Competing financial interests:** The authors declare no competing financial interests.

**Reprints and permission** information is available online at http://npg.nature.com/reprintsandpermissions/

**How to cite this article:** Hughes, M. A. *et al.* N-type chalcogenides by ion implantation. *Nat. Commun.* 5:5356 doi: 10.1038/ncomms6346 (2014).